\documentclass[aps,reprint,longbibliography,superscriptaddress]{revtex4-1}   %superscriptaddress (final style with ordered Authors)  aps,prl,reprint, groupedaddress

\usepackage{amsmath,amsfonts,amssymb}
\usepackage{wrapfig}
\usepackage{graphicx}
\usepackage{bbm}
\usepackage{float}
\usepackage{color}
\usepackage[unicode=true,colorlinks=true,citecolor=blue,urlcolor=blue]{hyperref}
\usepackage{braket}
\usepackage{bm}

\let\ifr\i

\renewcommand{\Re}{\mathop{\rm Re}\nolimits}
\newcommand{\e}{\mathrm{e}}
\renewcommand{\d}{\mathrm{d}}
\renewcommand{\i}{{\rm i}}

\newcommand{\tr}{{{\rm T}}}

\newcommand{\aver}[1]{\left\langle #1\right\rangle}

\begin{document}

\title{Theory of spin inertia in singly-charged quantum dots}

\author{D.~S.~Smirnov}
\affiliation{Ioffe Institute, 194021 St. Petersburg, Russia}
\author{E.~A.~Zhukov}
\affiliation{Experimentelle Physik 2, Technische Universit\"at Dortmund, 44221 Dortmund, Germany}
\author{E.~Kirstein}
\affiliation{Experimentelle Physik 2, Technische Universit\"at Dortmund, 44221 Dortmund, Germany}
\author{D.~R.~Yakovlev}
\affiliation{Experimentelle Physik 2, Technische Universit\"at Dortmund, 44221 Dortmund, Germany}
\affiliation{Ioffe Institute, 194021 St. Petersburg, Russia}
\author{D.~Reuter}
\affiliation{Department Physik, Universit\"at Paderborn, 33098 Paderborn, Germany}
\author{A.~D.~Wieck}
\affiliation{Angewandte Festk\"orperphysik, Ruhr-Universit\"at Bochum, 44780 Bochum, Germany}
\author{M.~Bayer}
\affiliation{Experimentelle Physik 2, Technische Universit\"at Dortmund, 44221 Dortmund, Germany}
\affiliation{Ioffe Institute, 194021 St. Petersburg, Russia}
\author{A.~Greilich}
\affiliation{Experimentelle Physik 2, Technische Universit\"at Dortmund, 44221 Dortmund, Germany}
\author{M.~M.~Glazov}
\affiliation{Ioffe Institute, 194021 St. Petersburg, Russia}

\begin{abstract}
The spin inertia measurement is a recently emerged tool to study slow spin dynamics, which is based on the excitation of the system by a train of circularly polarized pulses with alternating helicity. Motivated by the experimental results reported in E. A. Zhukov et al., arXiv:1806.11100 we develop the general theory of spin inertia of localized charge carriers. We demonstrate that the spin inertia measurement in longitudinal magnetic field allows one to determine the parameters of the spin dynamics of resident charge carriers and of photoexcited trions, such as the spin relaxation times, longitudinal $g$-factors, parameters of hyperfine interaction and nuclear spin correlation times.
\end{abstract}

\maketitle

\section{Introduction}

The spin dynamics of charge carriers in semiconductors has been a subject of intense studies during the few past decades. This is related to both fundamentally new physics~\cite{dyakonov_book,Bechtold2015} and possible future applications in semiconductor spintronics devices~\cite{Imamoglu1999}. The most interesting ones from the practical point of view are the systems with long spin relaxation and spin coherence times. The spin coherence of the free charge carriers in bulk semiconductors and semiconductor nanostructures is strongly quenched by, e.g., the Elliot-Yafet and Dyakonov-Perel spin relaxation mechanisms. By contrast, the spin relaxation of electrons and holes localized on impurities or in quantum dots can be very long and reach a few hundreds of nanoseconds in the absence of an external magnetic field and a few microseconds in moderate magnetic fields~\cite{KKavokin-review,eh_noise}.

To investigate slow spin dynamics, different approaches have been established. Among those are the Hanle effect, pump-probe techniques~\cite{glazov:review} and their  extensions~\cite{Extended_pp,dyakonov_book}, resonant spin amplification~\cite{Kikkawa98,yugova12}, and spin noise spectroscopy~\cite{Zapasskii:13,Oestreich-review,SinitsynReview}. Each of these approaches is aimed at the investigation of certain aspects of the spin dynamics in specific experimental conditions. Generally, all these techniques complement each other.

Surprisingly the most interesting question about the longitudinal spin relaxation time of localized charge carriers in an external magnetic field can be hardly addressed utilizing the approaches above. Only recently the spin inertia measurement has been suggested to measure spin relaxation times in the (sub-)microsecond range~\cite{Korenev2015}. This method is based on the excitation of the system by a train of circularly polarized pulses with alternating helicity. An increase of the helicity modulation frequency results in reduction of the spin polarization degree when this frequency exceeds the inverse spin relaxation time~\cite{PetrovBoxModel09,Ignatiev05,Fras2011}. This allows one to determine the longitudinal spin relaxation time in the structure under study.

The spin inertia is useful to study long spin relaxation times, which are known to be reached for localized charge carriers. The experimental results reported in Ref.~\onlinecite{SI_PRL} show the involved dependence of spin polarization on the magnitude of the external magnetic field and the modulation frequency. Moreover, this dependence turns out to be qualitatively different for $n$- and $p$-type quantum dots. This indicates the necessity of a detailed theoretical investigation of the spin inertia of localized charge carriers. In this paper we report on the results of such an investigation and demonstrate, that the spin inertia measurement gives access not only to the spin relaxation time, but also to the hyperfine interaction constants and longitudinal $g$-factors both of resident charge carriers and nonequilibrium electron-hole complexes.

The paper is organized as follows. In Sec.~\ref{sec:theory} we provide a general theoretical description of the spin inertia in terms of the Green's functions of spin dynamics. Then in Sec.~\ref{sec:ill} we consider two illustrative models of spin dynamics: monoexponential spin relaxation and precession in the Overhauser field of frozen nuclear spin fluctuations. The main Sec.~\ref{sec:main} is devoted to the detailed analysis of the most experimentally relevant model of spin dynamics in quantum dots, namely, the model which takes into account a finite nuclear spin correlation time. In the subsequent Sec.~\ref{sec:saturation} we briefly analyze the possible effect of spin polarization saturation. In Sec.~\ref{sec:compare} we compare the spin inertia method with the spin noise spectroscopy. Finally in Secs.~\ref{sec:discussion} and~\ref{sec:conclusion} we discuss the range of applicability of our theory and give concluding remarks, respectively.

\section{Theoretical description of spin inertia}
\label{sec:theory}

The resident carrier spin dynamics $\bm S(t)$ in spin inertia experiments can be described by a linear inhomogeneous equation of the form
\begin{equation}
  \frac{\d\bm S}{\d t}=\bm{\mathcal R}\left\lbrace\bm S\right\rbrace+\bm \Gamma(t).
  \label{eq:S_dyn}
\end{equation}
Here $\bm{\mathcal R}$ is a linear operator, which describes the spin evolution (precession and relaxation caused by the external magnetic field and the hyperfine interaction) between the pump pulses and does not explicitly depend on time, while $\bm\Gamma(t)$ describes the spin pumping rate. Equation~\eqref{eq:S_dyn} does not account for the saturation of electron spin polarization and implies that the spin polarization is much less than $100\%$. The effects of the polarization saturation are briefly discussed in Sec.~\ref{sec:saturation}.
Equation~\eqref{eq:S_dyn} can be formally solved introducing the dyadic Green's function $G_{\alpha\beta}(\tau)$ ($\alpha,\beta=x,y,z$) as
\begin{equation}
  S_\alpha(t)=\int\limits_{-\infty}^\infty G_{\alpha\beta}(\tau)\Gamma_\beta(t-\tau)\d\tau.
  \label{eq:S_G}
\end{equation}
We note, that the causality principle ensures that $G_{\alpha\beta}(\tau)=0$ for $\tau<0$.

We assume, that the pump pulses are elliptically or circularly polarized and arrive at time moments $t=K T_R$, where $K$ is an integer number and $T_R$ is the pump pulse repetition period. Therefore the spin generation rate can be represented as
\begin{equation}
  \Gamma(t)=\bm e_z\sum_{K}\Gamma_p(t-KT_R){P_K},
  \label{eq:Gamma}
\end{equation}
where $\bm e_z$ is the unit vector along $z$ direction, $\Gamma_p(t)$ describes the spin orientation rate by a single $\sigma^+$ pump pulse, and $P_K$ is the helicity of the $K$th pump pulse, $-1\leqslant P_K\leqslant 1$. In spin inertia measurements the polarization is modulated between $+1$ and $-1$ at frequency $\omega_m$.

In experiment, the $z$ component of the electron spin is detected synchronously with the pump pulses by the train of linearly polarized probe pulses arriving at time moments $t=K T_R+\tau_{d}$, where $\tau_{d}$ is the delay between the probe and pump pulses. The pulse duration $1.5$~ps is much shorter than all the other timescales in the system, and in particular, than the trion recombination time $\tau_0$, which, in turn, is much shorter, than $T_R$. The measured signal is thus defined, up to a common factor, as~\cite{Korenev2015}
\begin{equation}
  L=\frac{1}{M}\left|\sum_{K=1}^{M}S_z(K T_R+\tau_{d})\e^{\i\omega_m(K T_R+\tau_{d})}\right|.
  \label{eq:si}
\end{equation}
Here $M\gg 1$ is the macroscopic number of probe pulses used to accumulate the spin inertia signal. In experiment $M\sim10^7$.

The spin orientation of a resident charge carrier under resonant excitation of the trion in quantum dots is related to two processes: (i) almost instantaneous  (i.e., at the picosecond time scale of the pump pulse duration) excitation of the singlet trion state and (ii) recombination of the spin-polarized trion~\cite{yugova09,glazov:review}. Accordingly the generation rate can be presented as ($0\leqslant \tau< T_R$)
\begin{equation}
  \Gamma_p(\tau)=\Gamma_0\delta(\tau)+\frac{S_{\tr,z}(\tau)}{\tau_0},
  \label{eq:Gamma_p}
\end{equation}
where $\delta(\tau)$ represents the Dirac delta function, $\Gamma_0\ll 1$ is proportional to the pump pulse power, $\tau_0$ is the trion lifetime, and $\bm S_\tr(\tau)$ is the trion pseudospin. Note, that $\Gamma_0$ is dimensionless, while $\Gamma_0\delta(\tau)$ has the dimension of a rate, see Sec.~\ref{sec:saturation} for more details. We recall that the trion pseudospin is determined by the spin degree of freedom of an unpaired carrier: In $n$-type quantum dots the $T^-$ trion is excited and its $z$-pseudospin components $\pm 1/2$ correspond to the $\pm 3/2$ spin components of the heavy-hole; in $p$-type quantum dots the $T^+$ trion is excited and its pseudospin is equal to the electron-in-trion spin. Similarly we assume, that $\bm S$ refers to either the electron spin (for $n$-type quantum dots) or the heavy-hole pseudospin (for $p$-type quantum dots) in the ground state.

The trion spin dynamics after a single pulse is described by [cf. Eq.~\eqref{eq:S_dyn}]
\begin{equation}
  \frac{\d\bm S_\tr}{\d \tau}={\bm{\mathcal R}}^\tr\left\lbrace\bm S_\tr\right\rbrace-\frac{\bm S_\tr}{\tau_0}-\Gamma_0\bm e_z\delta(\tau).
\end{equation}
Here ${\bm{\mathcal R}}^\tr$ similarly to ${\bm{\mathcal R}}$ is a linear operator describing the trion spin precession and relaxation and we take into account, that due to the optical selection rules a circularly polarized pump pulse orients the resident charge carrier and trion spins in opposite directions~\cite{yugova09}. Introducing by analogy with Eq.~\eqref{eq:S_G} the Green's function $G^\tr_{\alpha\beta}(\tau)$ of the trion spin dynamics as a solution of
\[
\frac{\d\bm S_\tr}{\d \tau}={\bm{\mathcal R}}^\tr\left\lbrace\bm S_\tr\right\rbrace+\delta(\tau),
\]
one finds
\begin{equation}
  S_{\tr,z}(\tau)=-\Gamma_0 G^\tr_{z z}(\tau)\e^{-\tau/\tau_0},
\end{equation}
where we have explicitly separated the exponential trion recombination. In accordance with Eq.~\eqref{eq:Gamma_p} we arrive at
\begin{equation}
\label{Gamma:p:expl}
\Gamma_p(\tau) = \Gamma_0[\delta(\tau) - G^\tr_{zz}(\tau) e^{-\tau/\tau_0}],
\end{equation}
Substitution of Eq.~\eqref{Gamma:p:expl} along with Eqs.~\eqref{eq:S_G} and \eqref{eq:Gamma} in Eq.~\eqref{eq:si} gives the result
\begin{widetext}
\begin{equation}
  L=\frac{\Gamma_0}{M}\left|\sum_{K=1}^M\sum_{K'=-\infty}^\infty\e^{\i\omega_m(KT_R+\tau_d)} P_{K-K'}\left[ G_{zz}(K'T_R+\tau_d) - \int_0^\infty \frac{d\tau}{\tau_0} e^{-\tau/\tau_0} G_{zz}^\tr(\tau) G_{zz}(\tau_d + K'T_R - \tau) \right]\right|.
\label{eq:Lt}
\end{equation}
\end{widetext}
Here we have taken into account that $\tau_0$ is by far shorter than the timescales of electron and trion spin dynamics [$\tau_0\ll |G_{zz}(\tau)/G_{zz}'(\tau)|, |G^\tr_{zz}(\tau)/G^{\tr'}_{zz}(\tau)|$]. That is why in the calculation of the spin which returns from the trion the integration is extended up to $+\infty$. The expression in brackets can be recast as
\begin{equation}
\label{gen:1}
\mathcal G = \int_0^\infty \frac{d\tau}{\tau_0} e^{-\tau/\tau_0}  \left[G_{zz}(t) -  G_{zz}^\tr(\tau) G_{zz}(t- \tau)\right],
\end{equation}
with $t=K'T_R+\tau_d$. We make use of the fact that for short $\tau_0$ the relevant values of $\tau$ are effectively limited and the trion Green's function $G_{zz}^\tr(\tau)$ is close to $1$ while $|G_{zz}(t) - G_{zz}(t-\tau)|\ll|G_{zz}(t)|$. Separating in Eq.~\eqref{gen:1} the terms where just one small factor (either $G_{zz}(t) - G_{zz}(t-\tau)$ or $G^\tr_{zz}(\tau) -1$) is present and omitting the remaining terms we have
\begin{multline}
\label{gen:2}
\mathcal G = \int_0^\infty \frac{d\tau}{\tau_0} e^{-\tau/\tau_0}  \{G_{zz}(t) -  G_{zz}(t- \tau) \\
+ G_{zz}(t)[1-G_{zz}^\tr(\tau)] \}.
\end{multline}

Further we introduce the following Fourier transforms:
\begin{subequations}
\begin{equation}
P_K=\sum_{N=-\infty}^\infty\tilde P_N\e^{-\i N\omega_mKT_R},
\end{equation}
\begin{equation}
\label{eq:tildeG}
G_{zz}(\tau) = \int_{-\infty}^\infty \tilde G_{zz}(\omega) e^{-\i \omega \tau} \frac{\d\omega}{2\pi},
\end{equation}
\end{subequations}
and a similar expression for $\tilde G_{zz}^\tr(\tau)$. Noteworthy, the Fourier transform of the Green's function $\tilde G_{zz}(\omega)$ has the dimension of a time, while $G_{zz}(\tau)$ is dimensionless, the same is true for $\tilde G_{zz}^\tr(\omega)$ and $G_{zz}^\tr(\tau)$. For the modulation of polarization with a rectangular envelope, Fig.~\ref{fig:spin_dyn}(a), relevant for the experimental realizations, one has
\begin{equation}
  \tilde P_N=\frac{2\sin{(\pi N/2)}}{\pi N}.
\end{equation}
Assuming that the modulation $\omega_m$ and repetition $\omega_R=2\pi/T_R$ frequencies are incommensurable and taking the limit of $M\to \infty$ one has from Eq.~\eqref{eq:Lt} by virtue of Eq.~\eqref{gen:2}
\begin{equation}
  L=\frac{\Gamma_0}{T_R}\left|\tilde P_1\sum_{N=-\infty}^{+\infty}\tilde G_{zz}(\nu_N)\e^{-\i N\omega_R\tau_d}\left(Q-\frac{\i \nu_N\tau_0}{1-\i \nu_N\tau_0}\right)\right|,
\label{eq:Lw}
\end{equation}
with $\nu_N=\omega_m+N\omega_R$ and
\begin{equation}
  Q=\int\limits_0^\infty\frac{\d\tau}{\tau_0}\left[1-G_{zz}^\tr(\tau)\right]\e^{-\tau/\tau_0},
  \label{eq:Q_def}
\end{equation}
being trion spin flip probability during its lifetime. The two terms in brackets in Eq.~\eqref{eq:Lw} describe the efficiency of resident electron spin pumping. The spin generation is related to the difference of spin before pump pulse arrival and after trion recombination. This in turn can be caused either by trion spin relaxation or by resident spin relaxation during trion lifetime. Accordingly the first term, $Q$, describes the trion spin relaxation and the second term describes the resident spin relaxation on the timescale $t\gtrsim T_R$.

In realistic systems the trion spin relaxation is usually the dominant mechanism of spin pumping. Assuming additionally, that $T_R$ is shorter than the characteristic timescale of spin dynamics one can simplify Eq.~\eqref{eq:Lw} as
  \begin{equation}
    L=\frac{2\Gamma_0}{\pi T_R}\left|\tilde G_{zz}(\omega_m)Q\right|,
    \label{S0}
  \end{equation}
i.e. leave only the term with $N=0$. In agreement with the definition Eq.~\eqref{eq:si}, $L$ is dimensionless. This expression clearly shows, that the spin inertia is determined by the absolute value of the Green's function at the helicity modulation frequency.

For the further analysis it is also useful to introduce the effective spin relaxation time of a resident carrier as
\begin{equation}
  T_1=\left|\tilde G_{zz}(0)\right|,
  \label{eq:T1}
\end{equation}
which has the dimension of a time along with the Fourier transform of the Green's function $\tilde G_{zz}(\omega)$.
In general the spin inertia signal can be calculated after Eq.~\eqref{eq:Lw} if the Fourier transform of the Green's function is known.

\section{Illustrative examples of spin inertia}
\label{sec:ill}

In this section we consider two illustrative examples where the spin inertia signal can be calculated analytically providing qualitative picture of the phenomenon. The analysis below allows us to elucidate the role of various time scales in the quantum dot system.

\subsection{Model of monoexponential spin relaxation}
\label{sec:qualy}

Let us first consider the simplest case of monoexponential decay of the resident carrier and trion spin polarization, so that ${\bm{\mathcal R}}\left\lbrace S_z{\bm e_z}\right\rbrace=-S_z{\bm e_z}/\tau_s$, and ${\bm{\mathcal R}}^\tr\left\lbrace S_{z,\tr}{\bm e_z}\right\rbrace=-{S_{z,\tr}\bm e_z}/\tau_s^\tr$. Here $\tau_s$ and $\tau_s^{\tr}$ are the spin relaxation times of the resident carrier spin and of the photogenerated trion (pseudo-)spin, respectively. We note that these equations imply the absence of a transverse magnetic field, but in general allow for the presence of a longitudinal magnetic field. The relevant Green's functions have the form:
  \begin{subequations}
    \begin{equation}
      G_{zz}(\tau)=\exp(-\tau/\tau_s)\Theta(\tau),
      \label{G:exp}
    \end{equation}
    \begin{equation}
      G^\tr_{zz}(\tau)=\exp(-\tau/\tau_s^\tr)\Theta(\tau).
      \label{Gtr:exp}
    \end{equation}
  \end{subequations}
Here $\Theta(\tau)$ is the Heaviside step function being $1$ for $\tau\geqslant 0$ and zero otherwise.
Substitution of these expressions into Eq.~\eqref{eq:Lt} yields
\begin{equation}
  L=\frac{2\Gamma_0}{\pi}\left|\frac{\tau_0}{\tau_s^\tr}-\frac{\tau_0}{\tau_s}\right|\left|\frac{\e^{-\tau_{d}/\tau_s}}{1-(1+\i\omega_m T_R)\e^{-T_R/\tau_s}}\right|.
  \label{eq:si3}
\end{equation}
Here the first factor
\[
\frac{2\Gamma_0}{\pi}\left|\frac{\tau_0}{\tau_s^\tr}-\frac{\tau_0}{\tau_s}\right|
\]
describes the efficiency of resident carrier spin generation. It occurs due to an imbalance of the charge carrier and the trion spin relaxation. Indeed, the trion excitation leads to instantaneous spin polarization in the ground state, but then the opposite spin returns during the trion recombination, and partially compensates the spin polarization in the ground state. The shorter is the trion lifetime, the smaller is the deviation of the spin in the trion from the spin in the ground state, and the better it compensates the spin in the ground state after recombination~\cite{zhu07,glazov:review}.

Under the typical conditions $\tau_d, T_R \ll \tau_s$ we have
  \begin{equation}
    L=\frac{2\Gamma_0}{\pi T_R}\left|\frac{\tau_0}{\tau_s^\tr}-\frac{\tau_0}{\tau_s}\right|
\frac{\tau_s}{\sqrt{1+(\omega_m \tau_s)^2}},
    \label{eq:Korenev}
    \end{equation}
which coincides (up to a prefactor) with the result of Ref.~\onlinecite{Korenev2015}. This expression describes the dependence of the spin inertia signal on the spin relaxation time and the polarization modulation frequency. The longer is the spin relaxation time, $\tau_s$, the larger is the accumulated spin polarization. On the other hand, the higher is the modulation frequency, the smaller is the spin polarization, because the spin polarization injected at different time moments destructively interferes with itself. This dependence of the spin inertia signal allows for the experimental measurement of the spin relaxation time performing experiments with different $\omega_m$.

This simple analysis reveals the main features of the spin inertia signal: The longer is the spin relaxation time of the resident charge carrier the more spin is accumulated. By contrast, the longer is the trion spin relaxation time the less efficient is the electron spin polarization and the less spin is accumulated. The increase of modulation frequency leads to an effective suppression of the spin polarization and a decrease of the spin inertia signal.

The phenomenological spin relaxation introduced in this subsection does not allow one to analyze the dependence of the spin inertia signal on the magnitude of the external magnetic field. However, qualitatively an increase of the longitudinal magnetic field can lead to an increase of the spin relaxation time both of the resident carriers and the trions. In order to quantitatively analyze this effect we consider the microscopic mechanism of nuclei-induced spin relaxation.

\subsection{Model of ``frozen'' nuclear spin fluctuation}
\label{sec:frozen}

In quantum dot structures the charge carrier spin dynamics is usually non-Markovian, i.e., is not determined by a single spin relaxation time. It can be described within the central spin model where the electron or hole interacts with the host lattice nuclei~\cite{Urbaszek}. The hyperfine interaction for holes is an order of magnitude weaker, than for electrons~\cite{Grncharova-nuclei,PhysRevB.78.155329,PhysRevB.79.195440,Chekhovich_Hyperfine}, and, moreover, is strongly anisotropic. Nevertheless, the spin relaxation of holes at zero magnetic field and small magnetic fields ($\lesssim 1$~T) is known to be dominated by the hyperfine interaction in self-assembled QDs~\cite{PhysRevLett.102.146601,Godden2012,Fras2012,Urbaszek,eh_noise}.
% (see also Refs.~\cite{Fallahi2010,Fras2011,desfonds:172108,PhysRevLett.106.027402,Urbaszek,eh_noise,Prechtel2016}).
Within the quasi-classical approach the nuclear spins can be considered as frozen, in which case the electron spin rapidly (on the time-scale of several nanoseconds) looses two thirds of its initial orientation and then slowly relaxes to zero~\cite{merkulov02,PhysRevLett.88.186802,PhysRevLett.94.116601} with a phenomenological rate $1/\tau_s$.

\begin{figure}[t]
  \centering
  \includegraphics[width=0.9\linewidth]{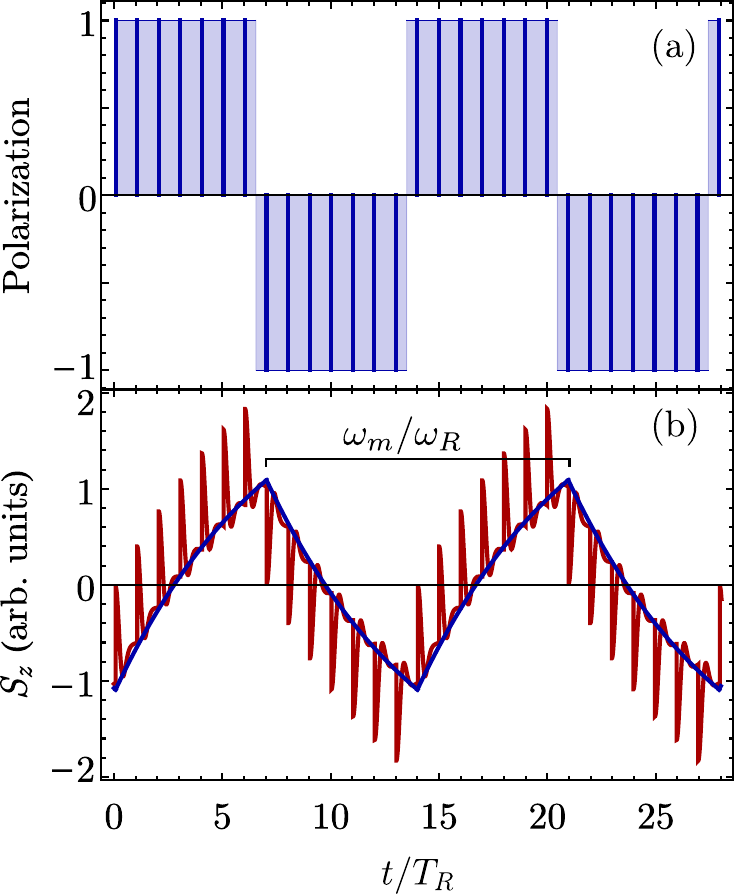}
  \caption{(a) Polarization of pump pulses $P_K=\pm1$ for $\omega_m=\omega_R/14$. The vertical lines indicate the pump pulse arrival moments. (b) Spin dynamics (brown curve) calculated after Eqs.~\eqref{eq:S_G} and~\eqref{eq:MER_G} with the parameters $\tau_s=100$~ns, $\omega_N=0.534$~ns$^{-1}$, $T_R=13.2$~ns. The blue curve shows the approximation with Eq.~\eqref{eq:MER_simple}. }
  \label{fig:spin_dyn}
\end{figure}

Figure~\ref{fig:spin_dyn}(a) illustrates the degree of pump pulse circular polarization~\footnote{In this illustrating figure commensurability between $\omega_m$ and $\omega_R$ is unimportant.}. The panel (b) shows the characteristic spin dynamics in the spin inertia experiment. For this calculation we have assumed that the trion spin dynamics is faster than the resident electron spin dynamics, so that one can neglect the second term in the brackets in Eq.~\eqref{eq:Lw}. In the calculations we took the electron Green's function in the form:
  \begin{multline}
    G_{\alpha\beta}(\tau)=\left\lbrace\frac{2}{3}\left[1-\frac{(\omega_N\tau)^2}{2}\right]\e^{-(\omega_N\tau)^2/4}+\frac{1}{3}\right\rbrace
    \\
    \times
    \e^{-\tau/\tau_s}\delta_{\alpha\beta}\Theta(\tau),
    \label{eq:MER_G}
  \end{multline}
which corresponds to the case of localized electrons, interacting with the static nuclear spin fluctuations~\cite{merkulov02}. Here $\omega_N$ is the characteristic spin precession frequency in the random Overhauser field, and $\tau_s$ is the spin relaxation time unrelated with the hyperfine interaction~\cite{PhysRevB.64.125316,PhysRevB.66.161318}. In order to simplify the qualitative analysis we limit ourselves in this subsection to the case of zero external magnetic field. One can see, that the spin dynamics is described by two timescales: The short timescale $\sim1/\omega_N$ ($\omega_N T_R \gg 1$) is characterized by the non-monotonic loss of $2/3$ of the injected spin polarization due to the dephasing in the random Overhauser field. The much longer timescale $\sim\tau_s$ is related with other mechanisms of spin relaxation. It is the longer timescale, which is important for the spin accumulation under the conditions of the spin inertia experiments.

It is seen from Fig.~\ref{fig:spin_dyn}(b) (brown curve) that the electron spin dynamics demonstrates the behavior characteristic to the central spin model with a fast non-monotonic decoherence followed by a slow relaxation, Eq.~\eqref{eq:MER_G}, see peaks in Fig.~\ref{fig:spin_dyn}(b), superimposed on smooth oscillations with the period being equal to the polarization modulation period, $2\pi/\omega_m$. This background, also shown by the blue curve in Fig.~\ref{fig:spin_dyn}(b), can be accurately described by considering the long time asymptotics of the Green's function~\eqref{eq:MER_G}, $\tau \omega_N \gg 1$, in the form:
  \begin{equation}
    G_{\alpha\beta}(\tau)=\frac{1}{3}\e^{-\tau/\tau_s}{\delta_{\alpha\beta}.}
    \label{eq:MER_simple}
  \end{equation}

\begin{figure}[t]
  \centering
  \includegraphics[width=0.9\linewidth]{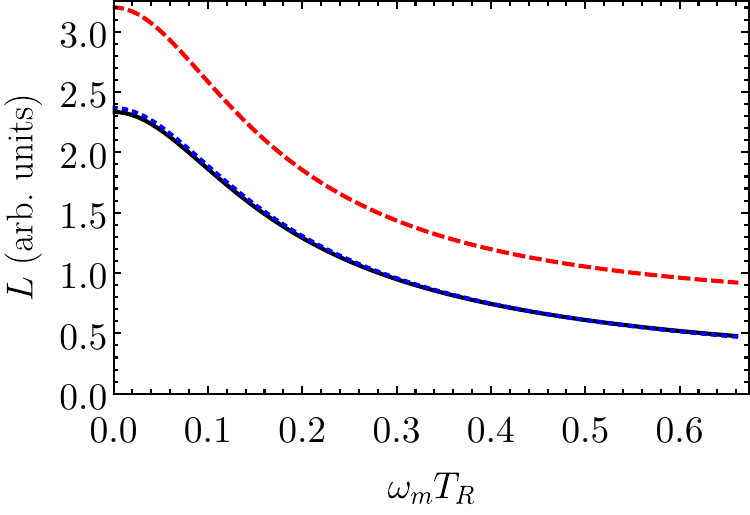}
  \caption{Spin inertia signal calculated after Eqs.~\eqref{eq:Lw} and~\eqref{eq:MER_G} for the same parameters as in Fig.~\ref{fig:spin_dyn} and $\tau_{d}=13$~ns (black solid curve) and $1$~ns (red dashed curve). The blue dotted curve shows calculation after simplified Eq.~\eqref{eq:Korenev}.}
  \label{fig:offset}
\end{figure}

Figure~\ref{fig:offset} shows the spin inertia signal calculated for two different pump-probe delays: $\tau_{d}\gg1/\omega_N$ (black curve) and $\tau_{d}\sim1/\omega_N$ (red curve). As expected the spin inertia signal decays with increasing $\omega_m$. However there is an offset between the two curves related to the fast timescales, when the spin dephasing in the random Overhauser field does not completely suppress $2/3$ of the injected spin polarization. This offset corresponds to the terms with $N\neq 0$ in the sum~\eqref{eq:Lw}. In fact, the case of $\tau_{d}\omega_N\gg1$ can be described by the simple approximation Eq.~\eqref{eq:Korenev} (blue dotted curve) due to the coincidence of the asymptotic form of the central spin model Green's function, Eq.~\eqref{eq:MER_simple}, and the Green's function describing the exponential spin relaxation, Eq.~\eqref{G:exp}. This is because in the spin accumulation at low modulation frequencies only the longitudinal (relative to the nuclear fluctuation) components of electron spin play a role.

In what follows we will assume that $\omega_N T_R\gg 1$, and that the pump-probe delay is close to $T_R$, which means that $T_R-\tau_{d}\ll T_R$. This analysis shows, that under the assumptions of fast trion spin relaxation (as compared to the ground spin relaxation) and not too high modulation frequencies $\omega_m\ll\omega_N$, one can leave only the term with $n=0$ in the general Eq.~\eqref{eq:Lw} for the spin inertia signal and use Eq.~\eqref{S0} to describe the experiments.

We note that in case of the monoexponential spin relaxation considered in Sec.~\ref{sec:qualy}, one simply has $Q=\tau_0/\tau_s^\tr$, and $T_1=\tau_s$. By contrast in the model of ``frozen'' nuclear spin fluctuations one has $T_1\approx\tau_s/3$, because only $1/3$ of the injected spin polarization contributes to the long timescale spin dynamics. We recall that the rate $1/\tau_s$ includes all spin relaxation mechanisms apart from hyperfine interaction. Using the same model for the trion spin dynamics one arrives at the expression for the spin generation rate:
  \begin{equation}
    Q=\frac{\tau_0}{\tau_s^\tr}+\left(\omega_N^\tr\tau_0\right)^2,
    \label{eq:Q_frozen}
  \end{equation}
where the superscript $\tr$ refers to the parameters of the trion spin dynamics.

\section{Spin dynamics accounting for finite correlation time of hyperfine fields}
\label{sec:main}

In self-assembled QDs the strain usually leads to the quadrupole splitting of the nuclear spin sublevels~\cite{Dzhioev2007,Urbaszek,Kuznetsova_quadrupole}, which induces nuclear spin dynamics. Indeed, the quadrupole splittings lead to random nuclear spin precession and changes of the Overhauser field orientation on the sub-microsecond timescale. The quadrupolar splittings of different nuclei interacting with the localized electrons or holes vary mainly due to the inhomogeneous strain in the quantum dot structures~\cite{PhysRevLett.109.166605}. Thus, the Overhauser field can be approximately characterized by the correlation time $\tau_c$~\footnote{We note that the numerical solutions of central spin model~\cite{Anders2016,Uhrig2017} can potentially yield more accurate description of the spin dynamics.}. The spin dynamics in this case can be described by the hopping model developed in Ref.~\onlinecite{Glazov_hopping}. Indeed the change of the nuclear spin configuration is equivalent to the charge carrier hopping between the different configurations of nuclear spins with rate $\tau_c^{-1}$. This approach allows us to investigate the electron spin dynamics in the presence of a longitudinal external magnetic field (Faraday geometry) accounting for the nuclear spin dynamics. As will be shown below a finite nuclear spin correlation time results in the increase of the spin inertia signal by more than three times with increase of the external magnetic field.

The spin in each nuclear spin configuration $i$ satisfies the equation:
\begin{equation}
  \frac{\d\bm S_i}{\d t}=\bm\Omega_i\times\bm S_i+\frac{1}{N}\sum_{j=1}^{N_c}\frac{\bm S_j-\bm S_i}{\tau_c}-\frac{\bm S_i}{\tau_s}.
  \label{eq:hopping}
\end{equation}
Here $\bm\Omega_i = \bm \Omega_{N,i}+\bm \Omega_L$ is the total electron spin precession frequency including both the Overhauser field in the given nuclear spin configuration ($\bm \Omega_{N,i}$) and the external field ($\bm \Omega_L$) contributions, $N_c$ is the number of the configurations. The distribution function of the Overhauser field has the form~\cite{merkulov02,NoiseGlazov,eh_noise}
\begin{equation}
  \mathcal F(\bm \Omega_N)=\frac{\lambda^2}{\left(\sqrt{\pi}\omega_N\right)^3}\exp\left(-\frac{\Omega_{N,x}^2+\Omega_{N,y}^2}{\omega_N^2/\lambda^2}-\frac{\Omega_{N,z}^2}{\omega_N^2}\right),
\end{equation}
where $\omega_N$ is the characteristic fluctuation of the Overhauser field along the growth direction, and $\lambda$ is the anisotropy parameter. For electrons $\lambda=1$ because their hyperfine interaction is isotropic and $\lambda>1$ for heavy holes~\cite{PhysRevLett.102.146601}. The effect of the external magnetic field on the nuclear spins can be neglected due to the very small nuclear magnetic moments.

The resident charge carrier spin is given by the  sum over all configurations:
\begin{equation}
  \bm S=\sum_{i=1}^{N_c}\bm S_i.
\end{equation}
Since the set of Eqs.~\eqref{eq:hopping} obeys the superposition principle, the total spin $\bm S(t)$ satisfies a linear equation of the form~\eqref{eq:S_dyn}, with $\bm{\mathcal R}$ being a non-local in time (non-Markovian) operator. Following Ref.~\onlinecite{Glazov_hopping} we present the corresponding Green's function in the presence of the longitudinal magnetic field $\bm B\parallel \bm e_z$ in the frequency domain as
\begin{equation}
  \tilde G_{zz}(\omega)=\frac{\tau_\omega\mathcal A}{1-\mathcal A\tau_\omega/\tau_c},
  \label{eq:Gzz}
\end{equation}
where $1/\tau_\omega=1/\tau_s+1/\tau_c-\i\omega$ and
\begin{equation}
  \mathcal A=\int d\bm \Omega_N \mathcal F(\bm \Omega_N)\frac{1+\Omega_{z}^2\tau_\omega^2}{1+\Omega^2\tau_\omega^2}.
  \label{eq:A}
\end{equation}
In the derivation of Eq.~\eqref{eq:A} we used the macroscopic limit ${N_c}\to \infty$ and replaced the summation over $\bm\Omega_{N,i}$ with the integration over $\bm\Omega_N$ with $\bm\Omega=\bm\Omega_L+\bm\Omega_N$.

Eqs.~\eqref{eq:Gzz}, \eqref{eq:A} can be also applied for the description of the trion spin dynamics provided that the appropriate distribution function $\mathcal F(\Omega)$ is chosen and in the definition of $\tau_\omega^{-1}$ the term $1/\tau_s$ is replaced by $1/\tau_0+1/\tau_s^{\tr}$. As a result one can show that accounting for the anisotropy of the hyperfine interaction and for the external magnetic field, Eq.~\eqref{eq:Q_frozen} should be modified as:
\begin{equation}
  Q={\frac{(\omega_N^\tr/\lambda^\tr)^2 \tau_0^2}{1+(\Omega_L^{\tr} \tau_0)^2}}+\frac{\tau_0}{\tau_s^{\tr}},
  \label{eq:Q}
\end{equation}
where $\Omega_L^{\tr}$ is the spin precession frequency of the unpaired charge carrier in the trion and the superscript $\tr$ is used for parameters of trion spin dynamics. One can see, that at short time scales the nuclear field correlation time does not affect the trion spin dynamics. The deviation of the trion spin from the $z$ direction during the trion lifetime is very small. In this case the longitudinal magnetic field hardly affects the spin dynamics, because, typically, $\Omega_L^{\tr} \tau_0$ is negligible. As a result only quite strong magnetic fields $B\sim\hbar/(|{g_{zz}^\tr}|\mu_B\tau_0)$ can suppress the trion spin precession in the Overhauser field. Even though the resident charge carrier spin relaxation is usually related to the hyperfine interaction on long timescales, the trion spin relaxation at short timescales can be dominated by the phenomenological time $\tau_s$.

In order to elucidate the role of the finite correlation time $\tau_c$ of the nuclear effective magnetic field it is instructive to qualitatively analyze the case of an isotropic hyperfine interaction, $\lambda=\lambda^\tr=1$. First let us discuss the dependence of $T_1$ on longitudinal magnetic field. At small magnetic fields $B\ll\hbar\omega_N/(|g|\mu_B)$ under the assumptions of $\omega_N\tau_c,\omega_N\tau_s\gg 1$ we arrive at the effective spin relaxation time, Eq.~\eqref{eq:T1},~\cite{Glazov_hopping}
\begin{equation}
  T_1=\frac{\tau_s}{3+2\tau_s/\tau_c}.
  \label{eq:T10}
\end{equation}
In case of an infinite nuclear spin correlation time, $\tau_c\to\infty$, the average spin relaxation time $T_1$ is three times shorter than the phenomenological time $\tau_s$ in agreement with the model of frozen nuclear spins presented in Sec.~\ref{sec:frozen} above~\footnote{We note that in case of holes the hyperfine interaction is anisotropic and $\lambda>1$. In this case more than $1/3$ of the initial spin polarization in conserved after spin dephasing in the random Overhauser field and $T_1>\tau_s/3$ in this limit.}. However, for the hyperfine field correlation time $\tau_c<\tau_s$ the Overhauser field changes direction during the electron spin lifetime, which leads to the additional suppression of spin polarization, as follows from Eq.~\eqref{eq:T10}. We note, that in case of $\tau_c\ll\tau_s$ one has $T_1=\tau_c/2$, which means, that the effective electron spin relaxation time is determined by the Overhauser field correlation time. This opens the way towards observation of the nuclear spin dynamics in the spin inertia signal.

\begin{figure}[t]
  \includegraphics[width=\linewidth]{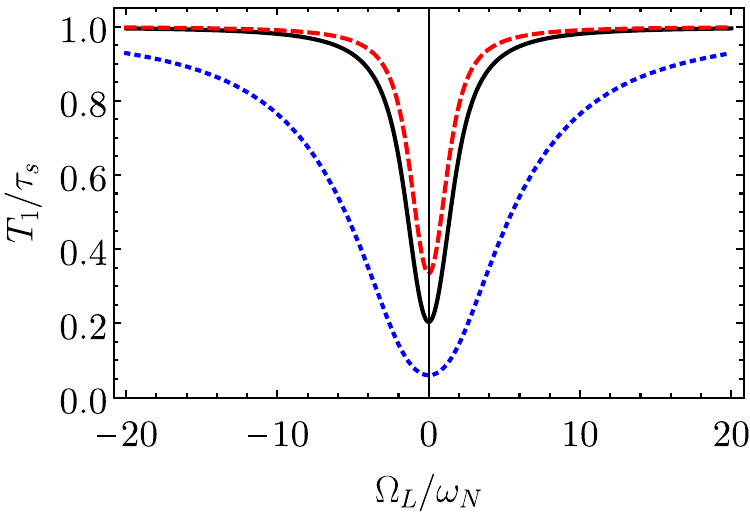}
  \caption{Effective spin relaxation time $T_1$ as a function of longitudinal magnetic field for different nuclear field correlation times: $\tau_c\omega_N=30$ (black solid line), $\infty$ (red dashed line), and $1$ (blue dotted line). The other parameters are the same as in Fig.~\ref{fig:trio}.}
  \label{fig:W}
\end{figure}

\begin{figure}[t]
\centering
\includegraphics[width=\linewidth]{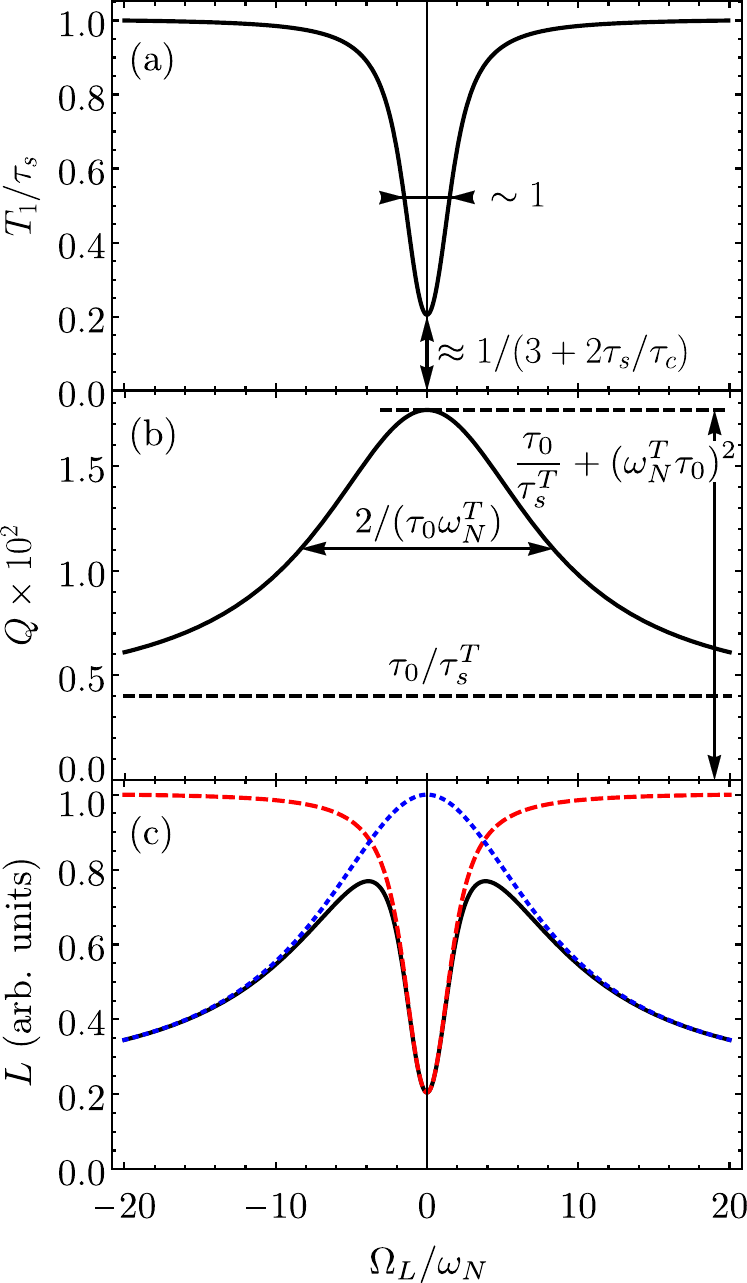}
\caption{(a) Effective spin relaxation time $T_1$ [Eq.~\eqref{eq:T1}] as a function of longitudinal magnetic field. (b) Relative spin generation rate $Q$ [Eq.~\eqref{eq:Q}] as a function of $B_z$. (c) Spin inertia signal $L$ (black solid line), calculated after Eq.~\eqref{S0} and~\eqref{eq:Gzz}, $T_1$ (red dashed line) and $Q$ (blue dotted line). The parameters of calculation are $\tau_c\omega_N=30$, $\tau_s\omega_N=30$, $\lambda=\lambda^\tr=1$, $\omega_N/g_{zz}=\omega_N^\tr/g_{zz}^\tr$, $\tau_0\omega_N^\tr=0.12$, $\tau_s\omega_N^\tr=30$, and $\omega_m=0$.
}
\label{fig:trio}
\end{figure}

The increase of spin polarization in the presence of a magnetic field is related to the suppression of spin dephasing in the Overhauser field. Provided that the Larmor precession frequency in the external field $\Omega_L=|g\mu_B B/\hbar|$ exceeds by far both the spin precession rate in the field of nuclear fluctuation, $\omega_N$, and the inverse correlation time, $\tau_c^{-1}$, one obtains for the effective spin relaxation time:
\begin{equation}
  T_1=\frac{\tau_s}{1+(\omega_N/\Omega_L)^2\tau_s/\tau_c}.
  \label{eq:T1:highB}
\end{equation}
It follows from Eq.~\eqref{eq:T1:highB} that $T_1$ increases with an increase of the external magnetic field. This dependence is shown in Fig.~\ref{fig:W}. For sufficiently large magnetic field the nuclear spin system becomes decoupled from the electrons and the effective spin relaxation time saturates at $\tau_s$. The amplitude of the dip at zero magnetic field depends on the Overhauser field correlation rate, as follows from Eq.~\eqref{eq:T1:highB} and is shown in Figs.~\ref{fig:W} and~\ref{fig:trio}(a). Interestingly in case of $\tau_c\omega_N\lesssim1$ the width of the dip increases with a decrease in the correlation time $\tau_c$. In this case instead of Eq.~\eqref{eq:T1:highB} one can use the following expression:
\begin{equation}
  \frac{1}{T_1}=\frac{1}{\tau_s}+\frac{\omega_N^2\tau_c}{1+(\Omega_L\tau_c)^2},
  \label{eq:T1:anyB:short}
\end{equation}
which is valid for an arbitrary relation between $\Omega_L$ and $1/\tau_c,\omega_N$ provided that $\omega_N\tau_c\ll 1$.

The Green's function, Eq.~\eqref{eq:Gzz}, allows one to calculate the spin inertia signal $L$ using Eqs.~\eqref{S0} and~\eqref{eq:Q}. Fig.~\ref{fig:trio}(c) shows the spin inertia signal for zero modulation frequency, calculated after Eq.~\eqref{S0}. The average spin relaxation time and spin generation rates are shown, respectively, in panels (a) and~(b). The characteristic parameters of these dependencies are indicated in the figure by arrows with labels. As before, we denote the parameters of the resident charge carrier spin dynamics $\omega_N,~\lambda,~\tau_c$ and $\tau_s$, and we add the superscript ``$\tr$'' for the same parameters in the trion spin dynamics. For example in case of a resident electron the singlet trion consists of a pair of electrons with antiparallel spins and an unpaired hole spin. In this case the parameters $\omega_N^\tr$ and $\lambda^\tr$ refer to the hole hyperfine interaction.

The spin polarization efficiency $Q$ given by Eq.~\eqref{eq:Q} is shown in Fig.~\ref{fig:trio}(b) as a function of $B$. This dependence is qualitatively similar to $T_1^{-1}$, but it is generally wider.
The spin inertia signal at small modulation frequencies $\omega_m \ll {1/T_1}$ is given by the product of the spin generation rate and spin relaxation time, $L\propto QT_1$~\footnote{The analysis of~Eqs.~\eqref{eq:Gzz} and~\eqref{eq:A} demonstrates that in the frequency range $\omega \ll \omega_N$ the Green's function in the frequency domain is given by $\tilde{G}_{zz}(\omega) = T_1/(1-\i \omega T_1/\mathcal A)$, where the frequency dependence of $\mathcal A$ is negligible. This results in the spin inertia signal of the form $L(\omega_m) \propto Q T_1/\sqrt{1+(\omega_m T_1/\mathcal A)^2}$.}. It is shown in Fig.~\ref{fig:trio}(c) by the black solid line. As follows from the above discussion the dependence of $L$ on magnetic field can be either monotonic, ``V-like'', provided that the polarization efficiency $Q$ weakly changes with variation of the magnetic field in the relevant field range or ``M-like''. Indeed, with the increase of magnetic field, first the spin inertia signal increases due to the suppression of the hyperfine induced spin dephasing and the increase of $T_1$. With further increase in the field $L$ goes down because of the suppression of the trion spin relaxation and the decrease of $Q$. We note that the latter effect takes place only when the hyperfine interaction is the main source of the trion spin relaxation, otherwise $Q$ does not depend on $B$. As follows from Eq.~\eqref{eq:Q} the latter situation is realized for $(\omega_N^{\tr})^2\tau_0\tau_s^\tr<1$.

\begin{figure}[t]
  \includegraphics[width=\linewidth]{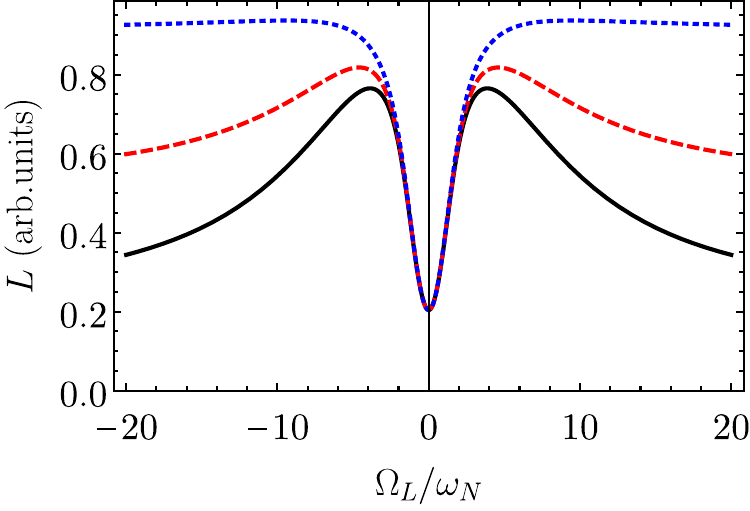}
  \caption{Spin inertia signal as a function of magnetic field for different trion spin relaxation times: $\tau_s^\tr\omega_N^\tr=30$ (black solid line), $8$ (red dashed line), and $1$ (blue dotted line). The other parameters including the modulation frequency and $\tau_0{\omega_N^\tr}=0.12$ are the same as in Fig.~\ref{fig:trio}.}
  \label{fig:tau_s}
\end{figure}

In Fig.~\ref{fig:tau_s} we show the modification of the spin inertia signal as a function of $B$ with a decrease of $\tau_s^\tr$. As follows from Eq.~\eqref{eq:Q} the shorter is the spin relaxation time, the less important is the nuclear induced spin relaxation. Therefore for short spin relaxation times the external magnetic field does not affect the trion spin relaxation and the resident charge carrier spin generation rate. As a result for very short $\tau_s^\tr$ (e.g. in case of $\tau_s^\tr{\omega_N^\tr\lesssim}1$) the dependence $L(B)$ becomes ``V-like''. We note, that this effect can take place with an increase of temperature provided $\tau_s^\tr$ is strongly temperature dependent, e.g. in case of phonon assisted spin relaxation~\cite{PhysRevB.64.125316,PhysRevB.66.161318}. Therefore the ``M-like'' shape of the spin inertia signal is a signature of hyperfine assisted trion spin relaxation, which leads to resident charge carrier spin polarization.

\begin{figure}[t!]
  \includegraphics[width=\linewidth]{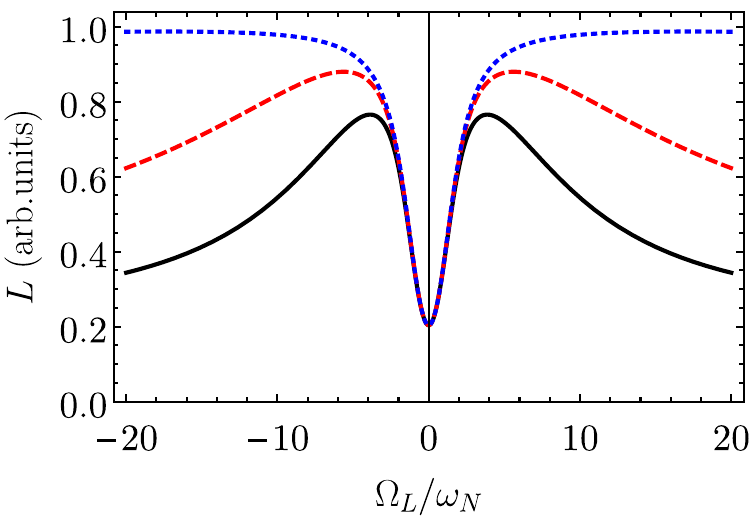}
  \caption{Spin inertia signal as a function of magnetic field for different trion recombination times: $\tau_0{\omega_N^\tr}=0.12$ (black solid line), $0.06$ (red dashed line), and $0.01$ (blue dotted line). The other parameters including $\omega_m=0$ and $\tau_s^\tr\omega_N^\tr=30$ are the same as in Fig.~\ref{fig:trio}.}
  \label{fig:tau_0}
\end{figure}

Interestingly, as it also follows from Eq.~\eqref{eq:Q} a similar effect takes place with decrease of $\tau_0$, as shown in Fig.~\ref{fig:tau_0}. Indeed at short timescales $\tau\ll1/{[(\omega_N^\tr)^2\tau_s^\tr]}$ the spin relaxation is always dominated by the trion spin relaxation (time constant is ${\tau_s^\tr}$), because the decrease of a trion spin projection $S_{z,\tr}$ due to the spin precession is quadratic in $\tau$. The main difference between the two dependencies illustrated in Figs.~\ref{fig:tau_s} and~\ref{fig:tau_0} is that a decrease of ${\tau_s^\tr}$ only makes the dependence $Q(B)$ flatter, while a decrease of $\tau_0$ also makes it wider, see labels in Fig.~\ref{fig:trio}(b).

\section{Polarization saturation effect}
\label{sec:saturation}

In this section we briefly analyze the effect of spin polarization saturation in case of strong pumping where the resident spin polarization may reach $100\%$. The account for the saturation effect requires modification of the spin generation rate, $\bm \Gamma(t)$ in Eq.~\eqref{eq:S_dyn} as compared with the simplified form used in Eq.~\eqref{eq:Gamma}. To that end we apply the quantum-mechanical model developed in Refs.~\cite{yugova09,spherical-dots}. It follows then that the spin components before, $S_\alpha^-$ ($\alpha=x,y,z$), and immediately after, $S_\alpha^+$, a single pump pulse accounting for the saturation effect are related as~\cite{yugova09,glazov:review,spherical-dots}
\begin{subequations}
  \begin{equation}
    S_z^+=P\frac{W}{4}+\left(1-\frac{W}{2}\right)S_z^-
  \end{equation}
  \begin{equation}
    S_x^+=WS_x^-,
    \quad
    S_y^+=WS_y^-,
  \end{equation}
\end{subequations}
where we have introduced the probability of trion excitation per pulse  $W=4\Gamma_0\in[0;1]$, and $P=\pm1$ stands for $\sigma^\mp$ polarized pump pulses. Here for the sake of simplicity we neglect the pulse induced spin rotation in the $(xy)$ plane assuming that the excitation pulses are exactly resonant with the trion excitation frequency. As a result, the spin generation term in Eq.~\eqref{eq:S_dyn} takes the form
\begin{multline}
  \bm \Gamma(t)=\sum_k\left\lbrace\frac{\bm e_z}{4}\left[P(t)-2S_z(t)\right]-\bm e_xS_x(t)-\bm e_y S_y(t)
    \right\rbrace
    \\\times
    4\Gamma_0\delta(t-KT_R)
    +\bm e_z \frac{S_{\tr,z}(t)}{\tau_0},
    \label{eq:Gamma_sat}
\end{multline}
where $\bm e_{x,y}$ are the unit vectors along the corresponding axes. We note that in the limit of small spin polarization one can neglect the terms, containing spin components, in Eq.~\eqref{eq:Gamma_sat} due to the fact that these contributions are at least quadratic in the pump power, which yields Eq.~\eqref{eq:Gamma} in the small pump power limit.

In order to qualitatively describe the effect of polarization saturation, we return to the model, described in Sec.~\ref{sec:qualy}, where the resident charge carrier and the trion spin dynamics are reduced to monoexponential spin relaxation. We also assume, that
\[
\tau_0\ll\tau_{d},T_R,\tau_s^\tr\ll\tau_s,\omega_m^{-1}.
\]
In this case the electron spin dynamics can be described by the continuous approximation, i.e., retaining only the first harmonic in modulation frequency both in Eq.~\eqref{eq:Gamma_sat} and in Eq.~\eqref{eq:Lw}:
\begin{equation}
  \frac{\d S_z}{\d t}=\frac{\Gamma_0\tau_0}{T_R\tau_s^\tr}\left[P(t)-2S_z\right]-\frac{S_z}{\tau_s}.
\end{equation}
From this equation one can see, that the polarization saturation simply results in the additive contribution to the effective spin relaxation rate proportional to the pump power~\footnote{The pumping also makes spin relaxation anisotropic~\cite{PhysRevB.89.081304}.}
\begin{equation}
\label{T1:saturation}
  \frac{1}{{\tau_s^*}}=\frac{1}{\tau_s}+\frac{2\Gamma_0\tau_0}{T_R\tau_s^\tr}.
\end{equation}
Accounting for this replacement $\tau_s\to {\tau_s^*}$, Eq.~\eqref{eq:Korenev} describes the spin inertia effect. In accordance with Eq.~\eqref{T1:saturation} the resident charge carrier spin relaxation time shortens with an increase in the pumping rate making the spin inertia signal dependence on $\omega_m$ flatter. Moreover, as expected, the electron spin polarization in this case is limited by $100\%$ since the product $|{\tau_s^*\Gamma_0}\tau_0/(T_R\tau_s^\tr)| \leqslant 1/2$.

\section{Comparison of spin inertia and spin noise measurements}
\label{sec:compare}

\begin{figure}[t]
 \includegraphics[width=\linewidth]{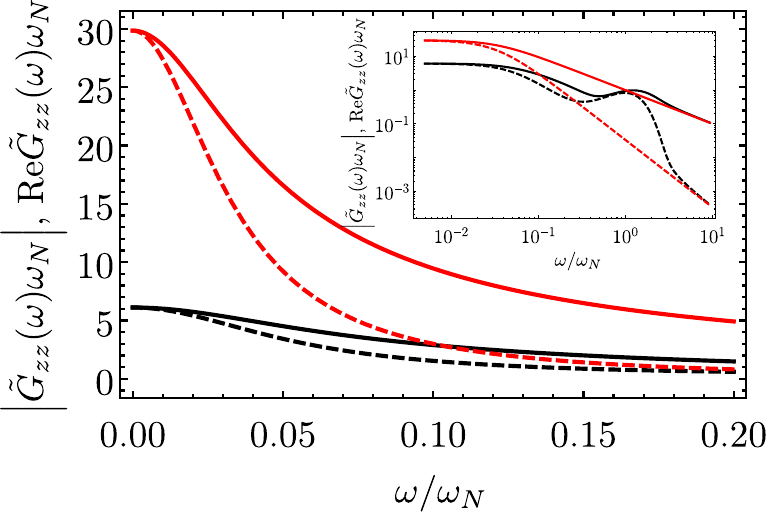}
  \caption{Modulus (solid lines) and real part (dashed lines) of the spin dynamics Green's function being proportional to the spin inertia signal and the spin noise spectrum, respectively. The parameters of calculation are the same as in Fig.~\ref{fig:trio}, and $\Omega_L=0$ for black curves, and $\Omega_L=20\omega_N$ for red curves.}
  \label{fig:compare}
\end{figure}

The spin inertia method is particularly convenient to study a long living spin polarization. Another experimental tool to study slow spin relaxation processes is the spin noise spectroscopy based on studies of the charge carriers spin correlation functions, $\langle \delta S_z(t)\delta S_z(t+\tau)\rangle$ in absence of external perturbation~\cite{Zapasskii:13,Oestreich-review,SinitsynReview}. It is instructive to compare these two recently emerged techniques of spin dynamics investigation. The spin fluctuations, $\delta S_z(t)$, obey, in equilibrium conditions, the same kinetic equation, Eq.~\eqref{eq:S_dyn},
as the optically induced spin polarization, with $\bm\Gamma(t)$ representing the fictitious Langevin forces,~\cite{ll5_eng}. Therefore the spin correlation function is determined by the same Green's function as:
\begin{equation}
  \aver{\delta S_z(t)\delta S_z(t+\tau)}\sim \tilde G_{zz}(\tau)+\tilde G_{zz}(-\tau).
\end{equation}
Here the angular brackets denote averaging over time $t$, and we took into account, that the correlator is an even function of $\tau$~\cite{ll5_eng}. The spin noise spectrum is defined as Fourier transform of the spin correlation function, therefore it is determined by:
\begin{equation}
  (\delta S_z^2)_\omega\sim\Re \tilde G_{zz}(\omega).
\end{equation}
This means, that while the spin inertia measurements provide the absolute value of the Green's function, see Eq.~\eqref{eq:T1}, the spin noise measurements yield its real part.

In order to clarify the difference we plot the spin noise spectrum and the spin inertia signals in Fig.~\ref{fig:compare}. One can see that the dependencies are qualitatively similar, but quantitatively different. Indeed, in general case, $\Re G_{zz}(\omega)\le|G_{zz}(\omega)|$. In particular, for exponential spin relaxation, the spin inertia signal decays as $1/\omega_m$ as a function of the modulation frequency, Eq.~\eqref{eq:Korenev}, while the spin noise spectrum falls as $1/\omega^2$~\cite{ll5_eng,NoiseGlazov,Glazov_hopping}. Hence, these two methods can complement each other. For instance, while measurements of the spin noise at very low frequencies are still very demanding, the spin inertia method can be applied there. By contrast, the spin noise spectroscopy can be effectively used for the frequencies $\omega\sim \omega_N$, which are hardly accessible by the inertia measurements.

\section{Discussion}
\label{sec:discussion}

We have shown that spin inertia measurements allow one to determine various parameters of the spin dynamics both of the charge carrier in the ground state and of the trion in the excited state. The method's time resolution is not limited by the repetition period $T_R \sim 13$~ns, and it makes possible to address long-living electron or hole spin coherence in quantum dots. It can also provide additional information beyond that extracted in the spin noise spectroscopy technique. Although the temperature does not directly affect the spin inertia signal, experiments performed at different temperatures can yield the temperature dependencies of the parameters of spin dynamics.

Despite the fact that the presented theory is aimed at a description of spin inertia in QD ensembles it can be also applied for impurity bound charge carriers and resident charge carriers localized at QW imperfections provided the hyperfine interaction dominates the resident carrier spin dynamics. In case of a free electron gas, the spin inertia signal can also provide information on spin relaxation times of resident charge carriers and trions. However, the hyperfine interaction, in this case, does not play a role.

Spin inertia measurements are most natural in Faraday geometry where the longest spin relaxation times can be reached. However, they can be useful to study any slow spin dynamics, for example in resonant spin amplification (RSA) conditions realized in the Voigt geometry~\cite{Kikkawa98,yugova12}. In this case, the transversal spin relaxation times can be accessed. In addition, a small deviation of the magnetic field in common spin inertia measurements from the $z$-direction can yield information about the transverse $g$ factor values, similar to Hanle measurements. To avoid any misunderstanding, we point out, that the ``M-like'' curves observed in Hanle experiments are caused by dynamic nuclear spin polarization~\cite{dyakonov_book}, and therefore have a different nature from the ``M-like'' curves predicted in this paper.

\section{Conclusion}
\label{sec:conclusion}

Spin inertia is a promising effect for measurement of the long longitudinal spin relaxation times in semiconductors. In this paper, we have studied theoretically the spin inertia of localized charge carriers. We have established the general theoretical description of spin inertia making use of dyadic Green's functions of the spin dynamics. We have considered a few experimentally relevant models of spin dynamics, and have shown that the capabilities of the spin inertia technique can go far beyond the measurement of the longitudinal spin relaxation times. In particular, we have demonstrated theoretically the opportunity to determine from the analysis of the spin inertia signal in a longitudinal magnetic field the $g$-factors and hyperfine interaction constants of the resident charge carriers, as well as of the resonantly excited trions. Depending on the structure parameters the dependence of the spin inertia signal on the longitudinal magnetic field can be either V- or M-like. The in-depth theoretical analysis performed in this paper has experimentally proved to be useful in Ref.~\onlinecite{SI_PRL}.

\begin{acknowledgments}
We acknowledge the financial support by the Deutsche Forschungsgemeinschaft in the frame of the International Collaborative Research Center TRR 160 (Project A5), partial support by the Russian Foundation for Basic Research
(Grant No. 15-52-12012), and Basis Foundation.
\end{acknowledgments}

\renewcommand{\i}{\ifr}
%\bibliography{Inertia_QDs_references}
%\bibliography{/home/dsmirnov/Practice/Theory/Lib/all-1.bib}


\end{document}